\documentclass[12pt,preprint]{aastex}

\slugcomment{Draft, version 13}

\shorttitle{[C~I] emission from the protoplanetary disk}
\shortauthors{T. Tsukagoshi et al.}

\usepackage{natbib}
\begin{document}


\title{First Detection of [C~I] $^3$P$_1$--$^3$P$_0$ Emission from a Protoplanetary Disk}


\author{%
Takashi TSUKAGOSHI\altaffilmark{1}
Munetake MOMOSE\altaffilmark{1}
Masao SAITO\altaffilmark{2}
Yoshimi KITAMURA\altaffilmark{3}
Yoshito SHIMAJIRI\altaffilmark{4}
and
Ryohei KAWABE\altaffilmark{5}
}

\altaffiltext{1}{College of Science, Ibaraki University, Bunkyo 2-1-1, Mito 310-8512, Japan: ttsuka@mx.ibaraki.ac.jp}
\altaffiltext{2}{Nobeyama Radio Observatory, Minamimaki, Minamisaku, Nagano 384-1305, Japan}
\altaffiltext{3}{Institute of Space and Astronautical Science, Japan Aerospace Exploration Agency, Yoshinodai 3-1-1, Sagamihara, Kanagawa 229-8510, Japan}
\altaffiltext{4}{Laboratoire AIM, CEA/DSM-CNRS-Universit${\rm \acute{e}}$ Paris, Diderot, IRFU/Service d'Astrophysique, CEA, Saclay,  F-91191 Gif-sur-Yvette Cedex, France}
\altaffiltext{5}{National Astronomical Observatory of Japan, Osawa 2-21-1, Mitaka, Tokyo 181-8588, Japan}

%\altaffiltext{3}{Joint ALMA Observatory, Alonso de Cordova 3107 OFC 129, Vitacura, Santiago 763 0355, Chile}


\begin{abstract}
We performed single point [C~I] $^3$P$_1$--$^3$P$_0$ and CO $J$=4--3 observations toward three T Tauri stars, DM Tau, LkCa 15, and TW Hya, using the Atacama Large Millimeter/submillimeter Array (ALMA) Band 8 qualification model receiver installed on the Atacama Submillimeter Telescope Experiment (ASTE).
Two protostars in the Taurus L1551 region, L1551 IRS 5 and HL Tau, were also observed.
We successfully detected [C~I] emission from the protoplanetary disk around DM Tau as well as the protostellar targets.
The spectral profile of the [C~I] emission from the protoplanetary disk is marginally single-peaked, suggesting that atomic carbon (C) extends toward the outermost disk.
The detected [C~I] emission is optically thin and the column densities of C are estimated to be $\lesssim10^{16}$ cm$^{-2}$ and $\sim10^{17}$ cm$^{-2}$ for the T Tauri star targets and the protostars, respectively.
We found a clear difference in the total mass ratio of C to dust, $M$(C)/$M$(dust), between the T Tauri stars and protostellar targets; the $M$(C)/$M$(dust) ratio of the T Tauri stars is one order of magnitude smaller than that of the protostars.
The decrease of the estimated $M$(C)/$M$(dust) ratios for the disk sources is consistent with a theoretical prediction that the atomic C can survive only in the near surface layer of the disk and C$^+$/C/CO transition occurs deeper into the disk midplane.
\end{abstract}

%In this paper, we report the first detection of [C~I] $^3$P$_1$--$^3$P$_0$ emission line from the protoplanetary disk around DM Tau.





\section{Introduction}
The gaseous component of protoplanetary disks is crucial for understanding the planet formation process because it affects the structures and evolution of the disk via chemical reactions.
Theoretical studies have predicted that protoplanetary disks have multiple layers in their vertical direction due to stellar radiation.
These include a relatively cool midplane, in which most of the molecules are depleted onto grains, a molecular-rich intermediate layer, and a hot surface layer \citep[e.g.,][]{bib:bergin2007}.
Because the physical environment differs among layers, gas phase abundances also vary significantly.\par

Atomic carbon (C) is considered to be abundant in the disk surface where far-ultraviolet (FUV) photons drive the energy balance and gas chemistry, i.e., the photon dominated region (PDR).
Theoretical studies predict that in the PDR, FUV photons with energies less than the ionization energy of hydrogen (13.6 eV) dissociate molecular hydrogen and carbon monoxide (CO) and ionize C to yield ionized carbon (C$^+$) \citep{bib:tielens1985}.
When significant attenuation is achieved, a thin H/H$_2$ transition layer should appear, beyond which  hydrogen molecules dominate.
Because the ionization energy of C is 11.6 eV, C$^+$ is dominant at the lower density regions of the PDR.
Deeper to the PDR, the carbon-ionizing radiation is attenuated, and CO forms via various chemical reactions.
As a result, for the uniform PDR, C exists in a thin layer sandwiched between the C$^+$ and CO layers.\par

Because photoevaporation at the disk surface is believed to govern the gas dispersal \citep{bib:hollenbach1994,bib:hollenbach2000}, observations of C are crucial for understanding disk dissipation.
Moreover, the observations of C are also important in terms of dust evolution in the disk because it affects the chemistry as a result of changes in the UV radiation field propagation and thus the intensity of the C line \citep{bib:jonkheid2004}.\par

However, detection of the submillimeter fine structure C lines has not been reported for protoplanetary disks thus far.
Although a few studies have attempted to detect the emission lines of C from protoplanetary disks around intermediate-mass stars, only upper limits have been obtained \citep{bib:chapillon2010,bib:panic2010,bib:casassus2013}.

In this paper, we present the results of [C~I] $^3$P$_1$--$^3$P$_0$ and $^{12}$CO $J$=4--3 observations with the Band 8 qualification model receiver mounted on the Atacama Submillimeter Telescope Experiment (ASTE)\footnote{The 10 m submillimeter telescope operated by Nobeyama Radio Observatory (NRO), a branch of National Astronomical Observatory of Japan, in collaboration with the University of Chile, and Japanese institutes including University of Tokyo, Nagoya University, Osaka Prefecture University, Ibaraki University, and Hokkaido University.}, and report the first detection of [C~I] emission from a protoplanetary disk.
We estimate the optical depth of the emission to derive the column density of C, and discuss possible evolutionary trends in the total mass of atomic C.

\section{Observations}
We performed [C~I] $^3$P$_1$--$^3$P$_0$ (492.161 GHz) and CO $J=$4--3 (461.041 GHz) observations toward three classical T Tauri stars (TTSs), \object{DM Tau}, \object{LkCa 15}, and \object{TW Hya}, in November 2010 using the ASTE.
Target information is listed in Table \ref{tab:info}.
In addition, we made [C~I] observations toward two protostars (PSs) in Taurus, L1551 IRS 5 and HL Tau, to check the consistency of the observational data.
We also used these data as a reference of the [C~I] intensity during the PS phase of star formation.\par

The observations were taken with an ALMA Band 8 (400--500 GHz) qualification model heterodyne receiver \citep[Band 8 QM;][]{bib:satou2008} mounted on the ASTE.
The Band 8 QM uses sideband separating receivers that detect two orthogonal polarizations and down-convert the sideband-separated intermediate frequency signals to 4--8 GHz.
The half power beam width (HPBW) was $17\arcsec$ and the main beam efficiency $\eta_\mathrm{MB}$ was estimated to be 45\% from observations of Jupiter.
For the backend, we used MAC, a 1024 channel digital auto-correlator, which has a band width of 128 MHz and a resolution of 125 kHz, corresponding to 78 and 0.076 km s$^{-1}$, respectively.

The position switch method was employed and a single spectrum toward the source position was obtained for each target.
Because our TTS targets are well known to be less affected by surrounding clouds, we selected OFF positions near the targets to effectively remove atmospheric fluctuations; 28, 15, and 1 arcmin from the stellar positions of DM Tau, LkCa 15, and TW Hya, respectively.
For the PSs, we employed an OFF position of ($\alpha_\mathrm{J2000}$, $\delta_\mathrm{J2000}$)=(4$^\mathrm{h}$ 32$^\mathrm{m}$ 6$\fs$0, 17$\arcdeg$ 48$\arcmin$ 51$\farcs$0), which was determined to avoid the CO $J=$1--0 emission of the L1551 cloud \citep{bib:yoshida2010}.

The telescope pointing calibration was performed every 1.5--2 hours by observing O-Cet and IRC+10216 in the CO $J=$4--3 emission line, and the resulting pointing accuracy was $\lesssim$3\arcsec.
Any time variation in intensity scale was checked by observing part of the Orion Horse-head nebula \citep[position A;][]{bib:philipp2006} two or three times daily and was found to be less than $\pm$10\%.
The [C~I] data were obtained under good sky conditions ($\tau_\mathrm{220GHz} \lesssim 0.05$), whereas $\tau_\mathrm{220GHz}=$0.06--0.10 for the CO observations.
The single sideband system noise temperatures were typically 2100 K and 3100 K for [C~I] and CO, respectively.

Data reduction and analysis were performed using the Common Astronomy Software Applications (CASA) version 3.3.0 software suite, in addition to its ASAP\footnote{ATNF Spectral Analysis Package} modules.
It should be noted that we reduced only one polarization dataset because of a beam alignment problem during the observations.
Bad data were removed by the {\it sdflag} task, baseline fitting and subtraction of the baseline from the spectra were performed with the {\it sdbaseline} task, and the data sets were combined in each source with the {\it sdaverage} task.
The final spectra were obtained after smoothing with a five pixel boxcar function along the channel axis by the {\it sdsmooth} task, corresponding to a velocity resolution of $\sim0.4$ km s$^{-1}$.\par

\begin{deluxetable}{ccccccccccccc}
\tablecaption{Target information}
\tabletypesize{\scriptsize}
\tablewidth{0pt}
\tablehead{
\colhead{Source} &
\colhead{R.A.} &
\colhead{Decl.} &
\colhead{$d$} &
\colhead{SP.} &
\colhead{$M_\ast$} &
\colhead{Age} &
\colhead{$M$(dust)} &
\colhead{$R_\mathrm{disk}$(CO)} &
\colhead{incl.} &
\colhead{$f_\mathrm{disk}$} &
\colhead{refs.}
\\
&
\colhead{(J2000)} &
\colhead{(J2000)} &
\colhead{(pc)} &
\colhead{type} &
\colhead{($M_\odot$)} &
\colhead{(Myr)} &
\colhead{(10$^{-4}$ $M_\sun$)} &
\colhead{(AU)} &
\colhead{($^\circ$)} &
\colhead{} &
\colhead{}
}
\startdata

DM Tau &
04:33:48.72 &
$+$18:10:10.0 &
140 &
M1 &
0.5 &
4.3 &
2.4 &
890 &
34 &
0.32 &
1,4,5,6\\

LkCa 15 &
04:39:17.80 &
$+$22:21:03.5 &
140 &
K5 &
1.0 &
3--12 &
4.8 &
905 &
52 &
0.25 &
1,6,7 \\

TW Hya &
11:01:51.91 &
$-$34:42:17.0 &
54 &
K7 &
0.8 &
10 &
3.4 &
215 &
7 &
0.15 &
2,8 \\
\hline
L1551 IRS5 & 04:31:34.07 & $+$18:08:04.9  & 140 & \nodata & \nodata  & \nodata & 5 & \nodata &\nodata &1 & 1\\
HL Tau & 04:31:38.44 & $+$18:13:57.7  & 140 & \nodata & \nodata & \nodata & 6 & \nodata & \nodata & 1& 1,3
\enddata
\tablerefs{(1) \citet{bib:andrews2005}; (2) \citet{bib:andrews2012}; (3) \citet{bib:beckwith1990}; (4) \citet{bib:kitamura2002}; (5) \citet{bib:oberg2010}; (6) \citet{bib:pietu2007}; (7) \citet{bib:qi2003}; (8) \citet{bib:wilner2000}}
\label{tab:info}
\end{deluxetable}

\section{Results}
\subsection{[C~I] $^3$P$_1$--$^3$P$_0$ and CO $J=$4--3 spectra}
\begin{figure}
   \plottwo{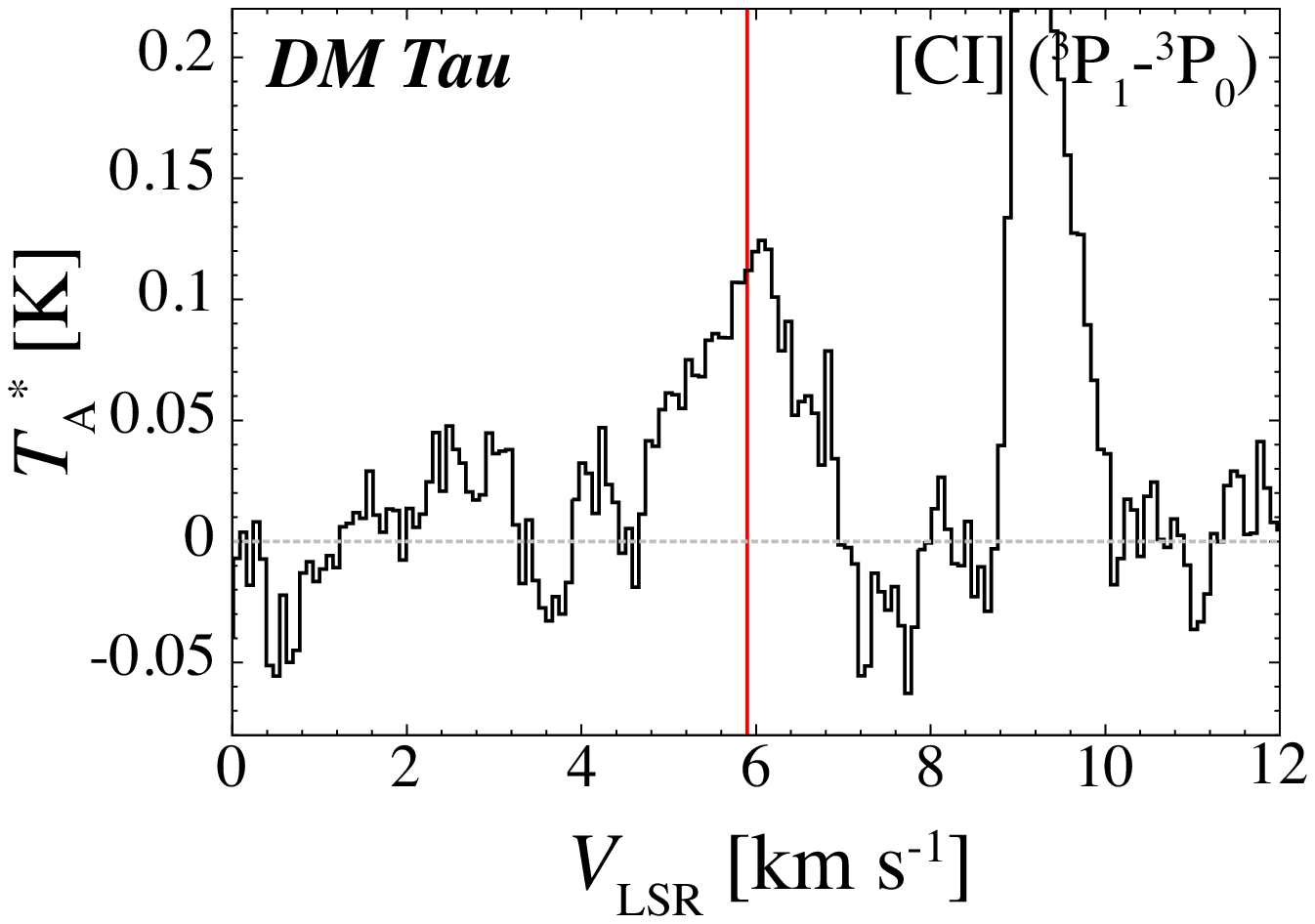}{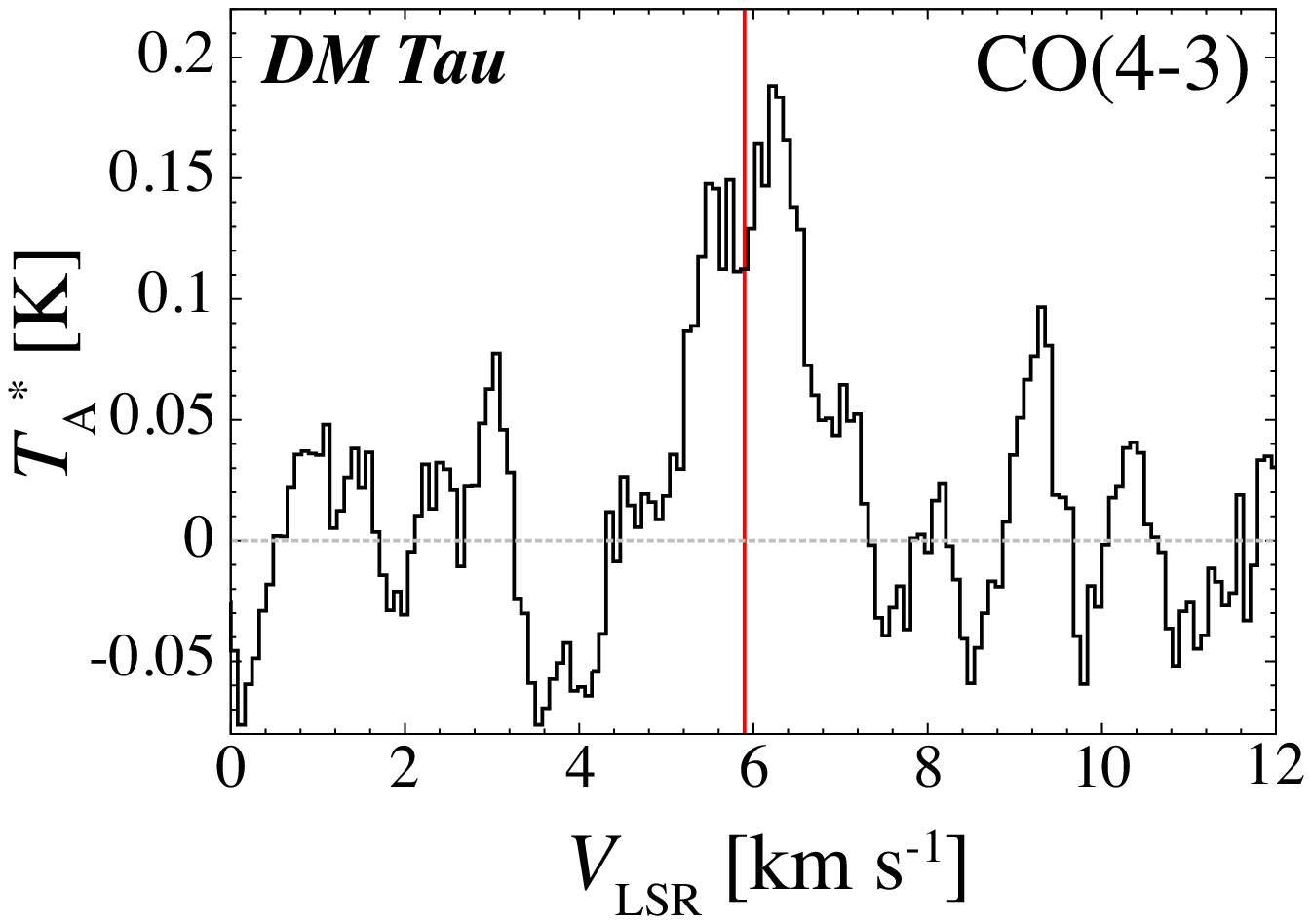}\\
   \plottwo{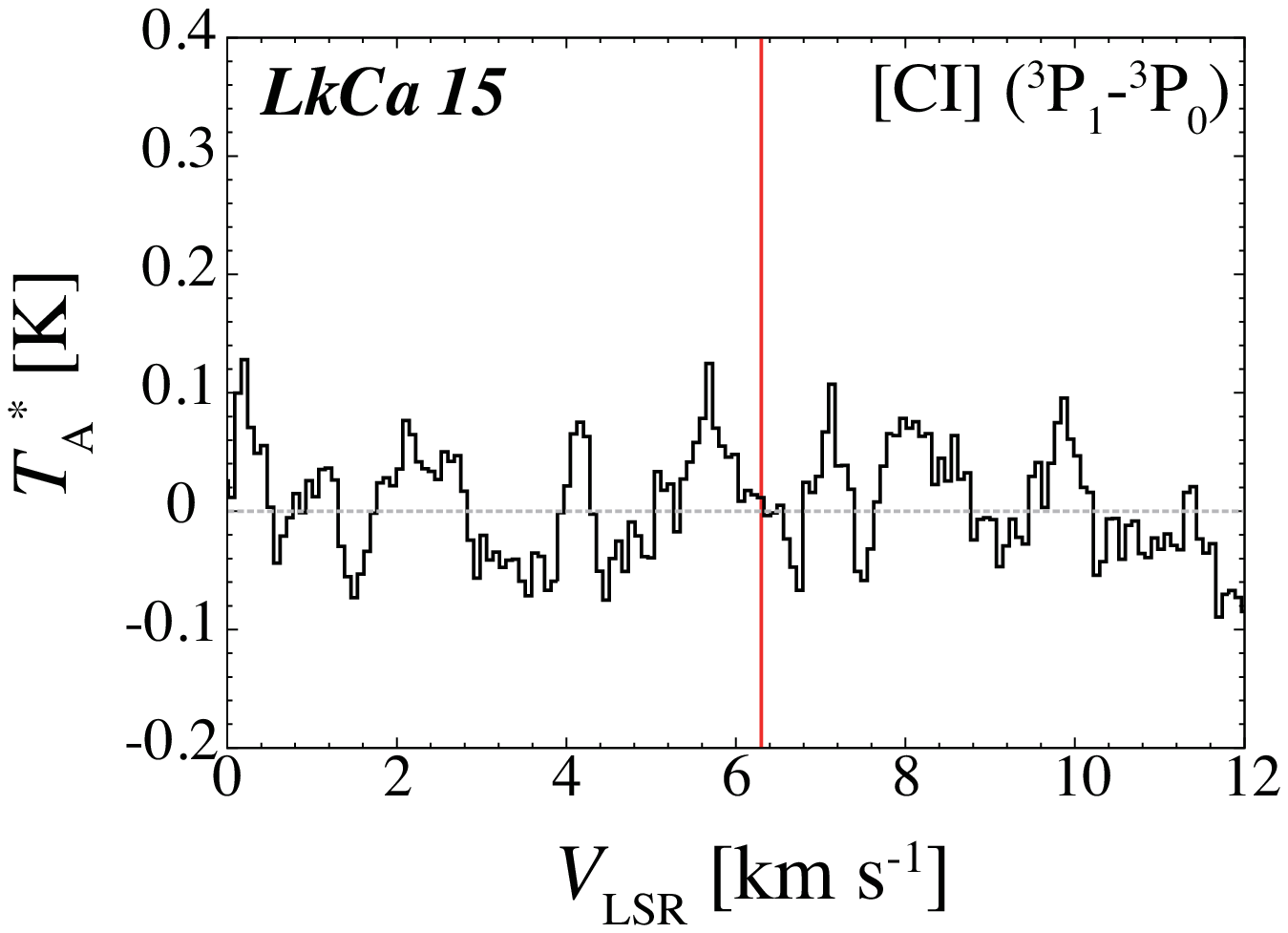}{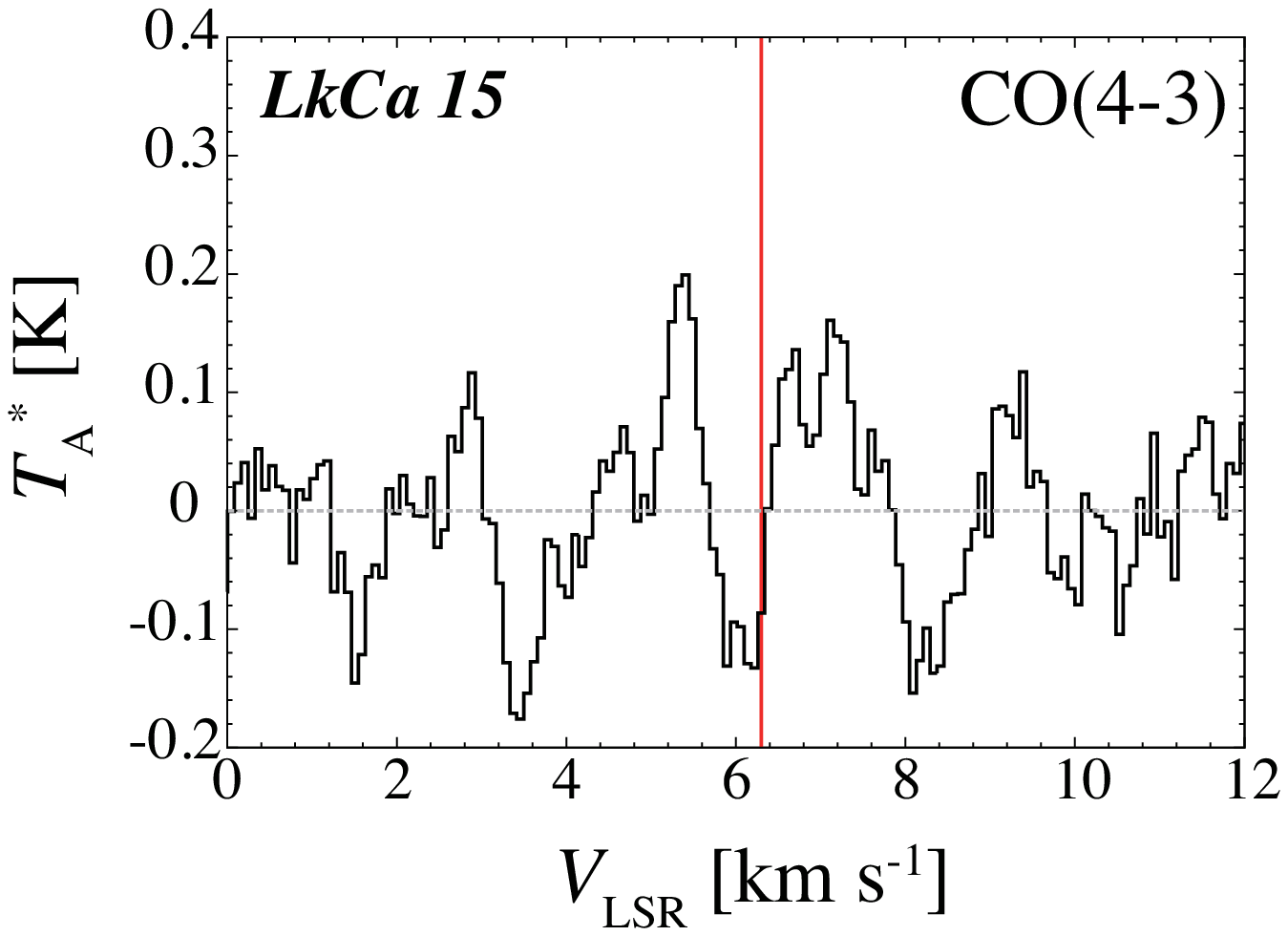}\\
   \plottwo{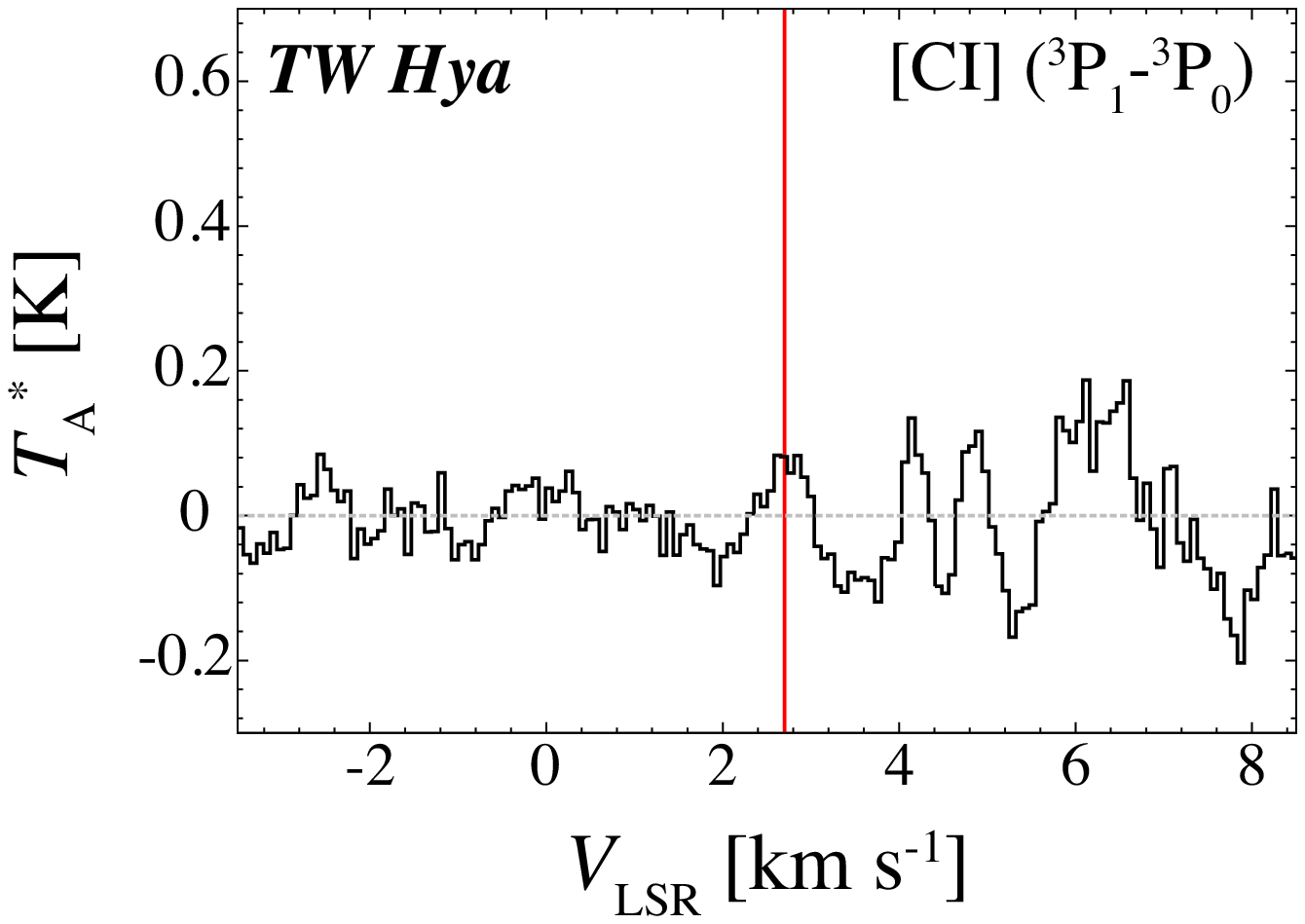}{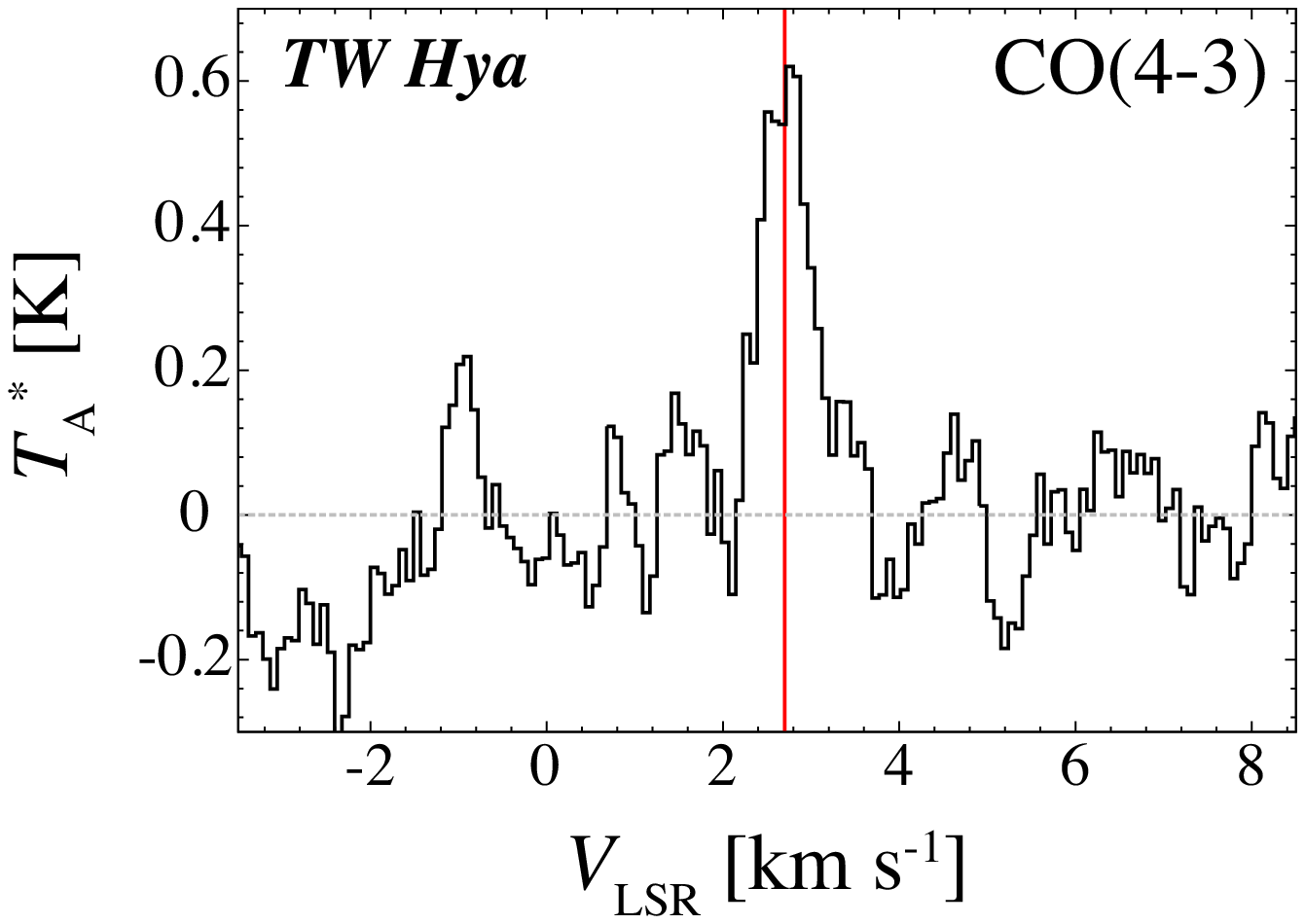}
   \caption{[C~I] (left) and CO (right) spectra toward the three TTSs, DM Tau (top), LkCa 15 (middle), and TW Hya (bottom). The vertical red line indicates the systemic velocity of the circumstellar disk \citep{bib:handa1995,bib:qi2003,bib:qi2004}.}\label{fig:spec_tts}
\end{figure}

Figure \ref{fig:spec_tts} shows the [C~I] $^3$P$_1$--$^3$P$_0$ and CO $J=$4--3 spectra toward the three T Tauri star targets, and the derived spectral line parameters are listed in Table \ref{tab:spec}.
We successfully detected [C~I] emission with a peak intensity of $T_\mathrm{A}^*=0.12$ K toward \object{DM Tau}.
The line profile is single peaked with a central velocity of $V_\mathrm{LSR}=$6.1 km s$^{-1}$, which is consistent with the systemic velocity of the circumstellar disk as estimated by molecular line observations \citep[5.9 km s$^{-1}$; e.g.,][]{bib:handa1995}.
The emission appeared to be slightly asymmetric with a somewhat larger extent to redshifted velocities.
This is similar to the spectral shape of the CO $J=$4--3  and other CO transitions. \citep{bib:guilloteau1994,bib:thi2001}.
Here, the emission component visible near $\sim9$ km s$^{-1}$ was not analyzed because it most likely originates from the ambient cloud \citep{bib:handa1995}.
For TW Hya and LkCa 15, we did not detect any [C~I] emission, and obtained only the 3$\sigma$ upper limits shown in Table \ref{tab:spec}.
To estimate upper limits to the total integrated intensity of the [C~I] emission lines, we assumed velocity widths of 3 and 1 km s$^{-1}$ for LkCa 15 and TW Hya, respectively, the values obtained by single dish CO observations \citep{bib:vanzadelhoff2001}.\par

\begin{figure}
    \plottwo{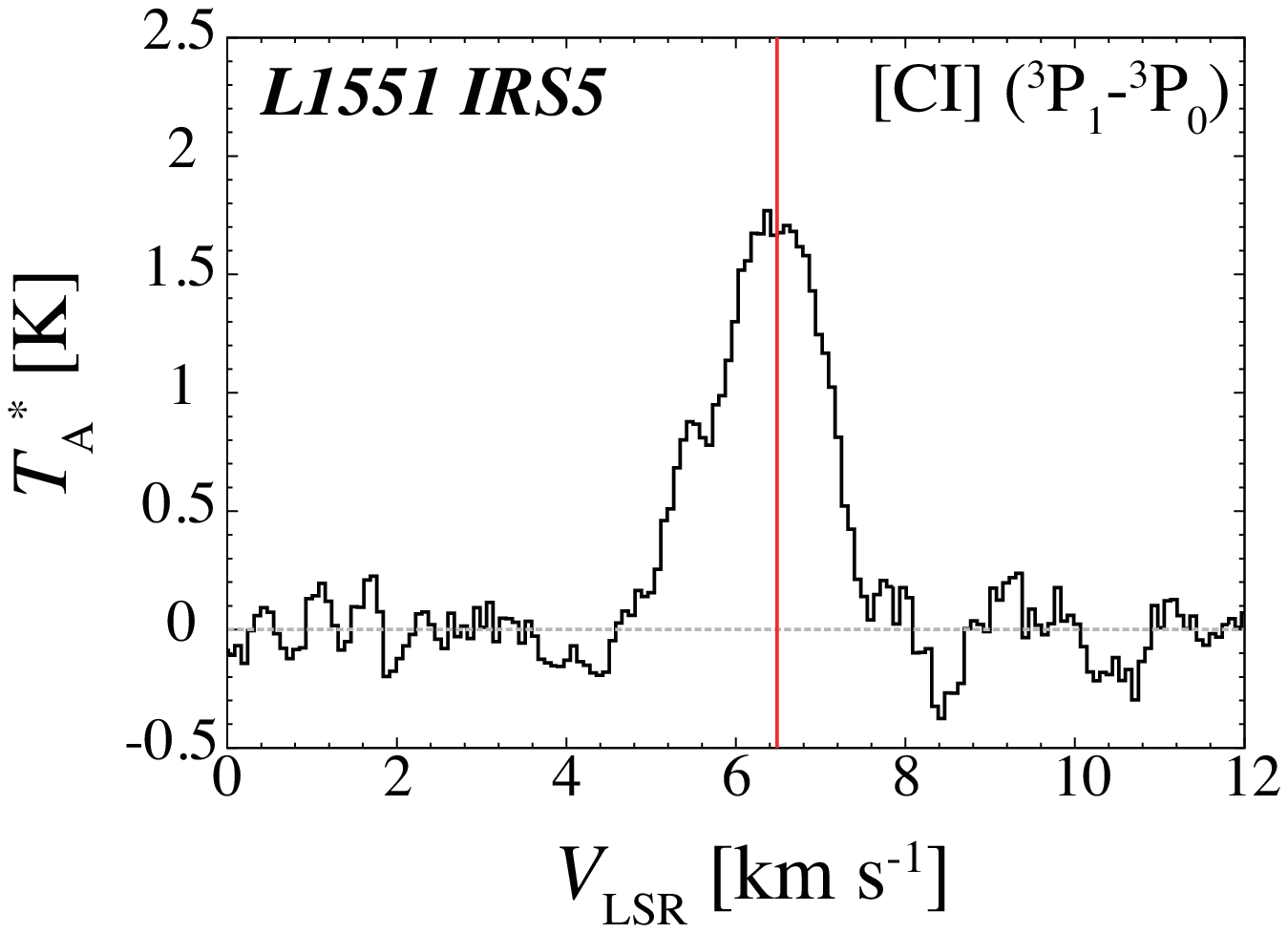}{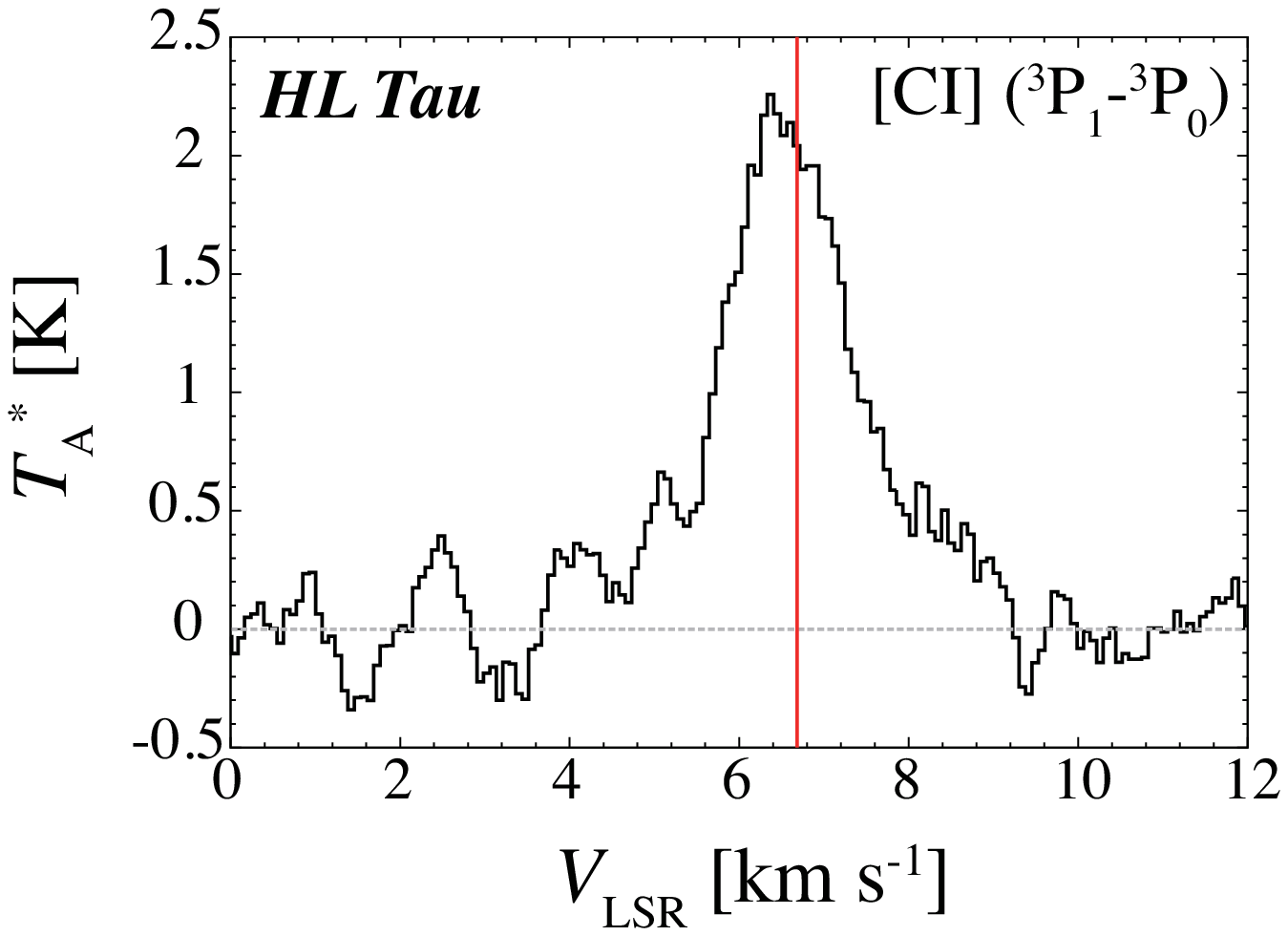}
    \caption{[C~I] spectra of the PSs, \mbox{L1551\ IRS5} (left) and \mbox{HL\ Tau} (right). The vertical red lines indicate the systemic velocity \citep{bib:saito2001}.}\label{fig:spec_ps}
\end{figure}

Significant [C~I] emission toward the PSs L1551 IRS 5 and HL Tau was detected, as shown in Figure \ref{fig:spec_ps}.
The spectral profiles are single-peaked at values close to their systemic velocities.
The emission intensities are markedly higher than that of DM Tau.\par

Emission in the CO $J$=4--3 line was detected toward \object{DM Tau} and \object{TW Hya}.
The peak velocities and line profiles are in good agreement with those of high-$J$ CO spectra \citep{bib:guilloteau1994,bib:thi2001,bib:vanzadelhoff2001}.
The double-peaked profile of DM Tau lies close to the systemic velocity.
After correcting for the beam size and efficiency, the integrated intensity of CO emission from TW Hya was comparable to the marginal value reported by \citet{bib:kastner1997} within the uncertainties.

\subsection{Optical depth of [C~I] emission and C column density}\label{sec:ci_mass}
We derived the optical depth of the [C~I] emission and thus the C column density using the equations reported by \citet{bib:oka2001}.
The optical depth at the peak velocity, $\tau_\mathrm{[C~I]}$, is given by
\begin{equation}
\tau_\mathrm{[C~I]} = -\ln \Biggr( 1- \frac{T_\mathrm{A}^\ast/\eta_\mathrm{MB}}{f_\mathrm{disk} [J(T_\mathrm{ex}) - J(T_\mathrm{bg}) ] } \Biggl) \mathrm{,}
\end{equation}
where $f_\mathrm{disk}$ is the beam filling factor of the source, $T_\mathrm{ex}$ is the excitation temperature, $T_\mathrm{bg}$ is the temperature of the cosmic background radiation, and $J$ is the radiation temperature defined by
\begin{equation}
J(T)=\frac{h\nu/k_\mathrm{B}}{\exp(h\nu/k_\mathrm{B}T)-1} \mathrm{.}
\end{equation}
Here $h$ is the Planck constant, $\nu$ is the observed frequency, and $k_\mathrm{B}$ is the Boltzmann constant.
The factor $f_\mathrm{disk}$ was derived from the ratio of the solid angle of the disk to that of the telescope's beam, $f_\mathrm{disk} =\Omega_\mathrm{disk}/\Omega_\mathrm{A}$.
The solid angle $\Omega_\mathrm{disk}$ was estimated by assuming that the extent of the [C~I] emission is identical to that of the resolved CO disk \citep{bib:pietu2007,bib:andrews2012}.
The outer radius of the CO disk and its inclination angle, in addition to the estimated $f_\mathrm{disk}$, are listed in Table \ref{tab:info}.
We employed $f_\mathrm{disk}=1$ for the PSs because they were expected to be embedded in extended envelopes that extend over the beam area.
For the TTSs, we assumed a constant  $T_\mathrm{ex}$ of 50 K, as suggested by theoretical studies of protoplanetary disks around typical T Tauri stars that predict the C$^+$/C/CO transition layer to appear at regions with $T_\mathrm{gas}\sim50$ K \citep{bib:jonkheid2004,bib:kamp2004}.
Conversely, we assumed $T_\mathrm{ex}=20$ K for the PSs, which is the accepted value for [C~I] observations of a PS surrounded by an infalling envelope \citep{bib:ceccarelli1998}.\par

Assuming that local thermodynamic equilibrium (LTE) applies, we derived the averaged C column density over the source via
\begin{eqnarray}
\nonumber N(\mathrm{C}) = 1.98\times10^{15} \frac{\int T_\mathrm{A}^\ast dv}{\eta_\mathrm{MB}} \ Q(T_\mathrm{ex}) \exp \Biggr( \frac{E_1}{k T_\mathrm{ex}} \Biggl) \\
\times \Biggr[ 1-\frac{J(T_\mathrm{bg})}{J(T_\mathrm{ex})} \Biggl]^{-1} \frac{\tau_\mathrm{[C~I]}}{1-\exp(-\tau_\mathrm{[C~I]})} f_\mathrm{disk}^{-1} \ \ \ \mathrm{.}
\end{eqnarray}
Here, $Q(T_\mathrm{ex})$ is the ground-state partition function for neutral atomic carbon,
\begin{equation}
Q(T_\mathrm{ex})=1+3 \exp \biggr( -\frac{E_1}{k T_\mathrm{ex}} \biggl) + 5 \exp \biggr( -\frac{E_2}{k T_\mathrm{ex}} \biggl) \ \ \ \mathrm{,}
\end{equation}
where $E_1$ and $E_2$ are the energies of the $J=$1 and 2 levels, respectively, and $\tau_\mathrm{[C~I]}$ is the velocity-averaged optical depth of the [C~I] emission.
In this study, however, we used the optical depth at the peak velocity instead of the velocity-averaged optical depth, which gives an upper limit to the column density.
It should be noted that we assumed the optically thin condition ($\tau_\mathrm{[C~I]}\ll1$) for LkCa 15 and TW Hya because no [C~I] emission was detected.
The C column density so estimated are listed in Table \ref{tab:quantity_ci}.
The total mass of atomic carbon, $M$(C), was derived by integrating the column density over the solid angle of the source.
Therefore, $M$(C) in the telescope beam can be uniquely determined regardless of source distribution.\par

Table \ref{tab:quantity_ci} clearly shows that all the detected [C~I] emission lines are optically thin.
The estimated C column density of the TTSs is typically $\sim10^{16}$ cm$^{-2}$ or less, whereas that of the PS is one order of magnitude higher ($\sim10^{17}$ cm$^{-2}$).
Accordingly, the $M$(C) values of the T Tauri star targets are one order of magnitude smaller than those of the PSs.
These results can be attributed to the lack of dense envelopes around the TTSs.\par

\section{Discussion}
\subsection{Origin of the [C~I] emission}
The [C~I] emission detected from DM Tau is most likely associated with the rotating gas disk around the star.
The peak velocity and full width half maximum are consistent with those of the disk in molecular lines \citep[e.g.,][]{bib:saito1995}.
Because of the modest upper energy of the ground state atomic C fine structure line ($\sim24$ K), the [C~I] emission is likely to be weighted toward the emission from the outer most part of the disk due to beam dilution.
In fact, the detected [C~I] emission shows a single-peaked profile near the systemic velocity, suggesting that the emission may dominate at the region where the rotation velocity is small, i.e., the outer part of the disk if Keplerian rotation applies.
In addition, the fact that the total mass of C is one order of magnitude smaller than those of the PSs (Table \ref{tab:quantity_ci}) supports the disk origin of the [C~I] emission.
However, the actual nature of the [C~I] emission of DM Tau remains unclear due to the large HPBW of the telescope.
High-resolution observations are required to reveal the detailed spatial distribution of the [C~I] emission over the disk.\par

The estimated [C~I] intensity of DM Tau is consistent with theoretical models of protoplanetary disk chemistry.
\citet{bib:jonkheid2004}, for example, calculated the [C~I] line intensity of a T Tauri star with a massive disk of 0.07 $M_\sun$ through a self-consistent treatment of the PDR near the disk surface.
The resultant peak intensity is expected to be $\sim$0.1 K in $T_\mathrm{A}^\ast$ after correcting for the ASTE beam size and the distance to the source, which is comparable to our observed intensity.\par

The higher intensities of the PSs than those of the TTSs suggest that the atomic carbon line originates mainly from the envelopes.
Because the velocity width of the [C~I] emission is much narrower than that of the wing emission of a molecular outflow ($\sim10$ km s$^{-1}$), the high velocity outflow is unlikely to affect the [C~I] emission \citep{bib:moriarty-schieven2006,bib:stojimirovi2006}.
There is a possibility of a low velocity cavity wall swept up by the outflow.
High-resolution [C~I] observations are required to separate the envelope from such a low velocity entrainment component.\par

\subsection{Difference of $M$(C)/$M$(dust) between T Tauri stars and protostars}\label{sec:4-2}
We introduce the total mass ratio of atomic C to dust grains, $M$(C)/$M$(dust)$ (\equiv R$), to investigate the evolutional variation of M(C).
The dust mass, which is generally estimated by optically thin (sub-)millimeter continuum observations, reflects the total gas mass of the system if the gas-to-dust ratio is constant.
Therefore, $R$ is an indicator of a fraction of C with respect to the total amount of gas, which is described as $R\propto g/d \times X\mathrm{(C)}$, where  $g/d$ is the gas-to-dust ratio, and $X\mathrm{(C)}$ is the fractional abundance of C with respect to H$_2$.
It is important to state that, especially for a protoplanetary disk, the atomic C and the dust grains trace distinct vertical layers of the disk; the atomic C presumably exists in the disk near-surface layers while the dust grains are mainly at the disk midplane.
Therefore, $R$ practically indicates the mass of C near the PDR with respect to the total mass.\par

We compiled $M$(dust) from previous studies, as listed in Table \ref{tab:info}.
All $M$(dust) data, except that for TW Hya, were calculated from the total masses in \citet{bib:andrews2005}, who measured (sub-)millimeter flux densities with single dish telescopes.
Their beam sizes were typically $\sim$10--15$\arcsec$, which is comparable to that of our [C~I] observations.
These authors estimated the total (gas+dust) disk mass by a simple disk model with power-law density and temperature profiles under an assumed gas-to-dust ratio of 100.
In Table \ref{tab:info}, we show the dust mass after correcting for the assumed gas-to-dust ratio.
For TW Hya, we took the mass estimate from \citet{bib:wilner2000}, who also used a similar simple disk model to estimate the disk mass.
Although their observations were made with an interferometer, the flux density of the entire disk was correctly recovered because the disk size is comparable to the detectable scale mentioned in \citet{bib:wilner2000}.\par

The derived $R$ is listed in Table \ref{tab:quantity_ci}.
A clear difference in $R$ is apparent between the TTSs and PSs; $R$ is typically an order of $10^{-4}$ or less for the TTSs, whereas the PSs show $R=3.4\times10^{-3}$, on average.
It should be noted that these results are not significantly affected by the difference in the assumed value of $T_\mathrm{ex}$.
If we use $T_\mathrm{ex}=$50 K for the PSs, which is the same value as in the TTSs, we obtain $R=2.7\times10^{-3}$, on average, a value still significantly higher than that for the TTS values.
When $T_\mathrm{ex}=$10 K, a typical value in molecular cloud cores, the difference was more noticeable.\par

As defined above, the decrement of $R$ can be interpreted by either a reduced $X\mathrm{(C)}$ or $g/d$.
The former case means that $X$(C) decreases as star formation progresses.
The chemical lifetime of C has been theoretically predicted to drop substantially with increments in the gas density \citep{bib:leung1984}.
Because the gas density significantly increases from a tenuous protostellar envelope to a dense disk near a TTS, a reduction in $X\mathrm{(C)}$ is expected at the TTS stage.
In fact, the typical density at the disk midplane is estimated to be $\sim10^9$ cm$^{-3}$ at 100 AU if we assume the minimum mass solar nebula \citep{bib:hayashi1981}, whereas the typical density of the PSs is $\sim10^{5}$ cm$^{-3}$ \citep{bib:momose1998,bib:saito2001}.
If the reduced $g/d$ is the case, the gas in the disk must be depleted to a low level of $g/d\lesssim10$ to explain the $R$ decrement of at least one order of magnitude.
However, such a high depletion of overall gas is not expected to occur in the disk, as discussed in \S \ref{sec:4-3}.\par

%However, such a highly depleted disk is at least inconsistent with the fact that the old TW Hya disk possess a large amount of gas \citep{bib:bergin2013}.\par

%\subsection{Difference among T Tauri stars}
\subsection{Non detection of [C I] emission toward LkCa 15 and TW Hya}\label{sec:4-3}
Only upper limits were obtained for the [C~I] emission from LkCa 15 and TW Hya, despite the fact that they also have disks as massive as that of DM Tau.
One possible explanation is that overall gas depletion with disk evolution leads to a depletion of C.
Since all the targets are in the evolutional stage of a so-called transitional disk, their disks are expected to be evolved with respect to a filled disk \citep{bib:andrews2011,bib:andrews2012}.
Moreover, their advanced stellar ages imply that the disks of LkCa 15 and TW Hya are older than the DM Tau disk.
The actual gas mass is difficult to directly estimate because of the large optical depths of molecular lines.
Some modeling efforts by using CO data indicate a lower gas-to-dust ratio with respect to the standard interstellar value \citep{bib:williams2014,bib:thi2010}.
However, as mentioned in \S \ref{sec:4-2}, the gas mass estimation for TW Hya by an HD line emission, which is believed to be a good tracer of gas mass, indicates that the disk is massive so that the gas-to-dust ratio is comparable to the standard interstellar value \citep{bib:bergin2013}.
Taking into account the fact that the disk model with a nominal gas-to-dust ratio of 100 reproduces well the observed intensity \citep{bib:jonkheid2004}, an overall gas depletion is unlikely the case.
An alternative is the depletion of volatiles such as C-containing species at the surface with disk evolution.
Such a time-dependent sink mechanism of C is proposed based on the observational data of HD and CO isotopologue lines \citep{bib:favre2013}.\par

\begin{deluxetable}{cccccccccc}
\tablecaption{Parameters of [C~I] $^3$P$_1$--$^3$P$_0$ and CO $J=$4--3 spectra}
\tabletypesize{\scriptsize}
\tablewidth{0pt}
\tablehead{
\colhead{} &
\multicolumn{4}{c|}{[C~I] $^3$P$_1$--$^3$P$_0$} &
\multicolumn{4}{c}{CO $J=$4--3}
\\  \cline{2-5}  \cline{6-9}
\colhead{Source} &
\colhead{$T_\mathrm{A}^\ast$ \tablenotemark{1}} &
\colhead{$V_\mathrm{LSR}$} &
\colhead{$\Delta V$} &
\colhead{$\int T_\mathrm{A}^\ast$ $dv$} &
\colhead{$T_\mathrm{A}^\ast$ \tablenotemark{1}} &
\colhead{$V_\mathrm{LSR}$} &
\colhead{$\Delta V$} &
\colhead{$\int T_\mathrm{A}^\ast$ $dv$} &
\\
\colhead{}&
\colhead{(K)} &
\colhead{(km s $^{-1}$)} &
\colhead{(km s $^{-1}$)} &
\colhead{(K km s $^{-1}$)} &
\colhead{(K)} &
\colhead{(km s $^{-1}$)} &
\colhead{(km s $^{-1}$)} &
\colhead{(K km s $^{-1}$)}
}
\startdata
DM Tau      & 0.12$\pm$0.03 & 6.1 & 1.4 & 0.17$\pm$0.05 & 0.18$\pm$0.03 & 6.2 & 1.1 & 0.20$\pm$0.05 \\
LkCa 15     & $<$0.15  & \nodata & \nodata & \nodata  &  $<$0.18 & \nodata & \nodata & \nodata \\
TW Hya      & $<$0.24  & \nodata & \nodata & \nodata & 0.62$\pm$0.08 & 2.8 & 0.7 & 0.43$\pm$0.11 \\
\hline
L1551 IRS 5 & 1.77$\pm$0.10 & 6.3 & 1.6 & 2.77$\pm$0.21 & \nodata & \nodata & \nodata & \nodata \\
HL Tau      & 2.26$\pm$0.08 & 6.4 & 2.0 & 4.55$\pm$0.18  & \nodata & \nodata & \nodata & \nodata \\
\enddata
\tablenotetext{1}{3$\sigma$ upper limit for non-detection.} 
\tablecomments{Column (1) gives the source name. Columns (2) and (6) give the peak intensity and root-mean-square noise level of the line in $T_\mathrm{A}^\ast$. Columns (3) and (7) give the local standard-of-rest (LSR) velocity at the emission peak. Columns (4) and (8) give the velocity width in full width at half maximum (FWHM) of the line. Columns (5) and (9) give the total integrated intensity of the line and its uncertainty.}
\label{tab:spec}
\end{deluxetable}

\begin{deluxetable}{cccccc}
\tablecaption{Physical parameters derived from [C~I] line.}
\tabletypesize{\scriptsize}
\tablewidth{0pt}
\tablehead{
\colhead{Source} &
\colhead{$\tau_\mathrm{[C~I]}$} &
\colhead{$N$(C)} &
\colhead{$M$(C)} &
\colhead{$\frac{M\mathrm{(C)}}{M\mathrm{(dust)}}$}
\\
\colhead{}&
\colhead{} &
\colhead{(10$^{16}$ cm$^{-2}$)} &
\colhead{(10$^{-7}$ $M_\sun$)} &
\colhead{(10$^{-4}$)}
}
\startdata
DM Tau      & 0.02 & 1.6$\pm$0.5 & 0.75$\pm$0.22 &  3.1 \\
LkCa 15     & $<$0.03 & $<$4.1 & $<$1.48 &  $<$3.1   \\ 
TW Hya      & $<$0.09 &  $<$3.6 & $<$0.12 &  $<$0.4  \\
\hline
L1551 IRS5      & 0.47 & 10.7$\pm$0.8 & 15.3$\pm$1.2 & 31  \\
HL Tau              & 0.65 & 19.0$\pm$0.8 & 27.4$\pm$1.1 & 46  \\
\enddata
\tablecomments{Column (1) gives the source name. Column (2) gives the optical depth of the [C~I] line. Column (3) gives the column density of C after correction for the beam filling factor of the disk. Column (4) gives the total mass of C. Column (5) gives the ratio of the total mass of C to the dust mass listed in Table \ref{tab:info}.}
\label{tab:quantity_ci}
\end{deluxetable}

%% The displaymath environment will produce the same sort of equation as
%% the equation environment, except that the equation will not be numbered
%% by LaTeX.


%% Included in this acknowledgments section are examples of the
%% AASTeX hypertext markup commands. Use \url without the optional [HREF]
%% argument when you want to print the url directly in the text. Otherwise,
%% use either \url or \anchor, with the HREF as the first argument and the
%% text to be printed in the second.

\acknowledgments
We acknowledge the ASTE staff for the operation and maintenance of the observational instruments.
We would like to thank H. Nomura and D. Ishimoto for fruitful discussions.
T.T. and M.M. were supported by JSPS KAKENHI grant No. 23103004.
Y.S. was supported by the French National Research Agency (grant No. ANR11BS560010 STARFICH).


%% in parentheses, after the keyword, but they will not be verified.

{\it Facilities:} \facility{Atacama Submillimeter Telescope Experiment}

\end{document}